\documentclass{SciPost}

\usepackage{amsmath, amssymb, amsthm}

\usepackage{graphicx}
\usepackage[export]{adjustbox}
\usepackage{hyperref}
\usepackage{microtype}
\usepackage{aas_macros}

\hypersetup{
    colorlinks,
    linkcolor={red!50!black},
    citecolor={blue!50!black},
    urlcolor={blue!80!black}
}

\usepackage[bitstream-charter]{mathdesign}
\DeclareSymbolFont{usualmathcal}{OMS}{cmsy}{m}{n}
\DeclareSymbolFontAlphabet{\mathcal}{usualmathcal}

\fancypagestyle{SPstyle}{
\fancyhf{}
\lhead{\colorbox{scipostblue}{\bf \color{white} ~SciPost Physics Lecture Notes }}
\rhead{{\bf \color{scipostdeepblue} ~Submission }}

\fancyfoot[C]{\textbf{\thepage}}
}

\newcommand{\dd}{\ensuremath{\text{d}}}
\newcommand{\ee}{\ensuremath{\text{e}}}
\newcommand{\ii}{\ensuremath{\text{i}}}
\renewcommand{\vec}[1]{\ensuremath{\boldsymbol{#1}}}
\newcommand{\bnabla}{\ensuremath{\boldsymbol{\nabla}}}
\newcommand{\mat}[1]{\ensuremath{\boldsymbol{\mathsf{#1}}}}
\newtheorem{theorem}{Theorem}
\newtheorem{definition}{Definition}

\begin{document}

\pagestyle{SPstyle}

\begin{center}{\Large \textbf{\color{scipostdeepblue}{
Bridging perturbation theory and simulations: initial conditions and fast integrators for cosmological simulations\\
}}}\end{center}

\begin{center}\textbf{
Oliver Hahn\textsuperscript{1,2$\star$},
}\end{center}

\begin{center}
{\bf 1} Dept. of Astrophysics, Univ. of Vienna, Türkenschanzstraße 17, 1180 Vienna, Austria
\\
{\bf 2} Dept. of Mathematics, Univ. of Vienna, Oskar-Morgenstern-Platz 1, 1090 Vienna, Austria
\\[\baselineskip]
$\star$ \href{mailto:oliver.hahn@univie.ac.at}{\small oliver.hahn@univie.ac.at}
\end{center}

\section*{\color{scipostdeepblue}{Abstract}}
\textbf{\boldmath{%
%
These lecture notes provide an introduction to the generation of initial conditions for cosmological $N$-body simulations. Starting from the definition and properties of Gaussian random fields, we discuss their role in cosmology and the efficient generation of such fields using Fourier methods. The Vlasov-Poisson system is introduced as the governing framework for cold collisionless matter, and its solution via characteristics and Lagrangian perturbation theory (LPT) is detailed. We discuss the use of LPT for initializing $N$-body simulations, emphasizing the importance of high-order LPT and late-time starts to minimize truncation and discreteness errors. Finally, we discuss time integration schemes, including PT-informed integrators, and their role in accurately evolving the system. These notes aim to bridge the gap between theoretical perturbation methods and practical simulation techniques.
}}

\vspace{\baselineskip}

\noindent\textcolor{white!90!black}{%
\fbox{\parbox{0.975\linewidth}{%
\textcolor{white!40!black}{\begin{tabular}{lr}%
  \begin{minipage}{0.6\textwidth}%
    {\small Copyright attribution to authors. \newline
    This work is a submission to SciPost Physics Lecture Notes. \newline
    License information to appear upon publication. \newline
    Publication information to appear upon publication.}
  \end{minipage} & \begin{minipage}{0.4\textwidth}
    {\small Received Date \newline Accepted Date \newline Published Date}%
  \end{minipage}
\end{tabular}}
}}
}


\vspace{10pt}
\noindent\rule{\textwidth}{1pt}
\tableofcontents
\noindent\rule{\textwidth}{1pt}
\vspace{10pt}


\title{Les Houches Summer School 2025 `The Dark Universe':\\ Bridging the gap between cosmological perturbation theory and simulations}
\author{O.Hahn}
\date{\today}

\section{Simulating Gaussian random fields}
\subsection{Definitions and properties}
Gaussian random fields (GRFs,\cite{Adler:2010}) are the starting point for all cosmological simulations as they are (1) very simple objects to work with, and (2) very well motivated by observations of the CMB (which is extremely close to a GRF although deviations from perfect Gaussianity are much sought for as hints of inflationary physics). 
\begin{definition}[Gaussian random field, GRF]
  A Gaussian random field (GRF) $f$ is a family of random variables parameterized over a domain $\mathcal{D}\subset\mathbb{R}^d$ such that for any finite set of points $\vec{x}_1,\ldots,\vec{x}_m\in\mathcal{D}$, the random variables $f(\vec{x}_1),\ldots,f(\vec{x}_m)$ are jointly Gaussian distributed. The joint distribution is fully specified by the mean vector $\mu(\vec{x}_i)$ and the covariance matrix $(\mat{C})_{ij}=C(\vec{x}_i,\vec{x}_j)$, i.e.
  \begin{align}
    \mathbb{E}[f(\vec{x}_i)] &= \mu(\vec{x}_i)\;, & \mathbb{E}[f(\vec{x}_i)f(\vec{x}_j)] &= C(\vec{x}_i,\vec{x}_j) + \mu(\vec{x}_i)\,\mu(\vec{x}_j)\;. 
  \end{align}
  We call $f$ \textbf{centered} if $\mu(\vec{x})= 0\;\forall\vec{x}$  (which can always be achieved by re-defining the field as $f-\mu$). For a centered field, we call $f$ \textbf{homogeneous} (or stationary) if the covariance is translation-invariant, i.e. if a function $\mathscr{C}$ exists so that 
  \begin{align}
    C(\vec{x}_i,\vec{x}_j) = \mathscr{C}(\vec{x}_i-\vec{x}_j)
  \end{align}
  and additionally \textbf{isotropic} if it is also rotation-invariant, i.e. if a $\xi$ exists so that
  \begin{align}
    C(\vec{x}_i,\vec{x}_j) = \xi(\|\vec{x}_i-\vec{x}_j\|)\;.
  \end{align}
\end{definition}
\begin{theorem}[Isserlis/Wick]
  Let $(f_1,\dots,f_m):=(f(\vec{x}_1),\dots,f(\vec{x}_m))$ be an $m$-variate Gaussian random variable with \emph{zero mean} (i.e. centered), then 
  \begin{align}
    \mathbb{E}[f_1\,f_2\,\cdots\,f_m] = \left\{
      \begin{array}{ll} 
      \sum_{p\in P_m^2}\;\prod_{\{i,j\}\in p} \underbrace{\mathbb{E}[f_i\,f_j]}_{=(\mat{C})_{ij}} & \textrm{ if $m$ is even} \\
      0 & \textrm{ if $m$ is odd}
      \end{array}
      \right. \;,
  \end{align}
  where the sum is over all pairings $p\in P_m^2$, the set of all ways to partition $\{1,\dots,m\}$ into pairs $\{i,j\}$, and the product is over the pairs in $p$. The proof can be found in most textbooks on statistics. 
\end{theorem}
\noindent The meaning of the theorem is that the expectation value of the product of an even number of centered Gaussian random variables can always be expressed in terms of the covariance matrix $\mat{C}$, while the expectation value of an odd number of centered Gaussian random variables vanishes. I.e. for a Gaussian field, the three-point correlator vanishes and the four-point correlator can be expressed in terms of two-point correlators, and so on. Conversely, for a non-Gaussian field, this is not true and can be used to quantify deviations from Gaussianity (e.g. through a non-vanishing three-point correlator).

\subsection{Diagonality of covariance in Fourier space, sampling}
A GRF is a GRF irrespective of the basis in which it is represented. In particular, by linearity, the Fourier transform of a GRF is also a GRF, i.e. 
  \begin{align}
    \hat{f}(\vec{k}) &= \int_{\mathbb{R}^d} \dd^d x\; \ee^{-\ii \vec{k}\cdot\vec{x}} \, f(\vec{x})\;, 
  \end{align}
  is a complex GRF (where both the real and the imaginary part are GRFs).
\begin{theorem}[Diagonality in Fourier space]
  The Fourier-space covariance of a homogeneous centered field is diagonal with
  \begin{align}
    \mathbb{E}[\hat{f}(\vec{k})\hat{f}^*(\vec{k'})] &= (2\pi)^d \delta_D(\vec{k}-\vec{k'}) \mathscr{P}(\vec{k})\;, \label{eq:def_power}
  \end{align} 
  where $\delta_D$ is the Dirac-$\delta$ distribution, and $\mathscr{P}(\vec{k})$ is the power spectrum, defined as the Fourier transform of the covariance function 
  \begin{align}
    \mathscr{P}(\vec{k}) &:= \int_{\mathbb{R}^d} \dd^d r \; \ee^{-\ii \vec{k} \cdot \vec{r}} \mathscr{C}(\vec{r})\;.
  \end{align}  
  If the GRF is isotropic, then a $P$ exists so that $\mathscr{P}(\vec{k})=P(\|\vec{k}\|)$. Note that for real valued fields equivalently $\mathbb{E}[\hat{f}(\vec{k})\hat{f}(\vec{k'})] = (2\pi)^d \delta_D(\vec{k}+\vec{k'}) \mathscr{P}(\vec{k})$ since the Hermitian property holds, i.e. $\hat{f}^\ast(\vec{k})=\hat{f}(-\vec{k})$.
\end{theorem}
\begin{proof}
  To prove the theorem, the covariance function in Fourier space is defined as:
  \begin{align*}
  \mathbb{E}[\hat{f}(\vec{k}) \hat{f}^*(\vec{k'})] &= \int_{\mathbb{R}^d} \dd^d x \int_{\mathbb{R}^d} \dd^d y \; \ee^{-\ii (\vec{k} \cdot \vec{x} - \vec{k'} \cdot \vec{y})} \mathbb{E}[f(\vec{x}) f(\vec{y})].
  \intertext{
    Using the definition of the covariance function in real space, $\mathbb{E}[f(\vec{x}) f(\vec{y})] = C(\vec{x}, \vec{y})$, and substituting $C(\vec{x}, \vec{y}) = \mathscr{C}(\vec{x} - \vec{y})$ for homogeneous GRFs, we rewrite
  }
   &= \int_{\mathbb{R}^d} \dd^d x \int_{\mathbb{R}^d} \dd^d y \; \ee^{-\ii (\vec{k} \cdot \vec{x} - \vec{k'} \cdot \vec{y})} \mathscr{C}(\vec{x} - \vec{y}).
  \intertext{Changing variables to $\vec{r} = \vec{x} - \vec{y}$ and $\vec{X} = \vec{x}$, we have $\dd^d x \dd^d y = \dd^d X \dd^d r$ due to the unit Jacobian of this transformation. The integral then becomes}
  &= \int_{\mathbb{R}^d} \dd^d X \int_{\mathbb{R}^d} \dd^d r \; \ee^{-\ii (\vec{k} - \vec{k'}) \cdot \vec{X}} \ee^{-\ii \vec{k} \cdot \vec{r}} \mathscr{C}(\vec{r}).
  \intertext{
  and the integral over $\vec{X}$ yields a Dirac delta distribution
  $
  \int_{\mathbb{R}^d} \dd^d X \; \ee^{-\ii (\vec{k} - \vec{k'}) \cdot \vec{X}} = (2\pi)^d \delta_D(\vec{k} - \vec{k'}).
  $
  Substituting this, we finally obtain the key result}
  \mathbb{E}[\hat{f}(\vec{k}) \hat{f}^*(\vec{k'})] & = (2\pi)^d \delta_D(\vec{k} - \vec{k'}) \int_{\mathbb{R}^d} \dd^d r \; \ee^{-\ii \vec{k} \cdot \vec{r}} \mathscr{C}(\vec{r}).
  \end{align*}
  This completes the proof. The remaining integral is the Fourier transform of the covariance function $\mathscr{C}(\vec{r})$, which by definition \eqref{eq:def_power} equals the power spectrum $\mathscr{P}(\vec{k})$.
\end{proof}

\begin{figure}
  \centering
  \includegraphics[width=0.45\columnwidth,valign=c]{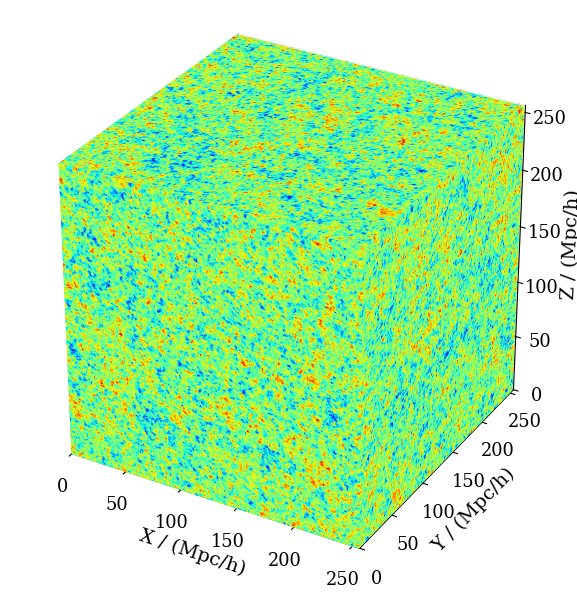}\qquad
  \includegraphics[width=0.5\columnwidth,valign=c]{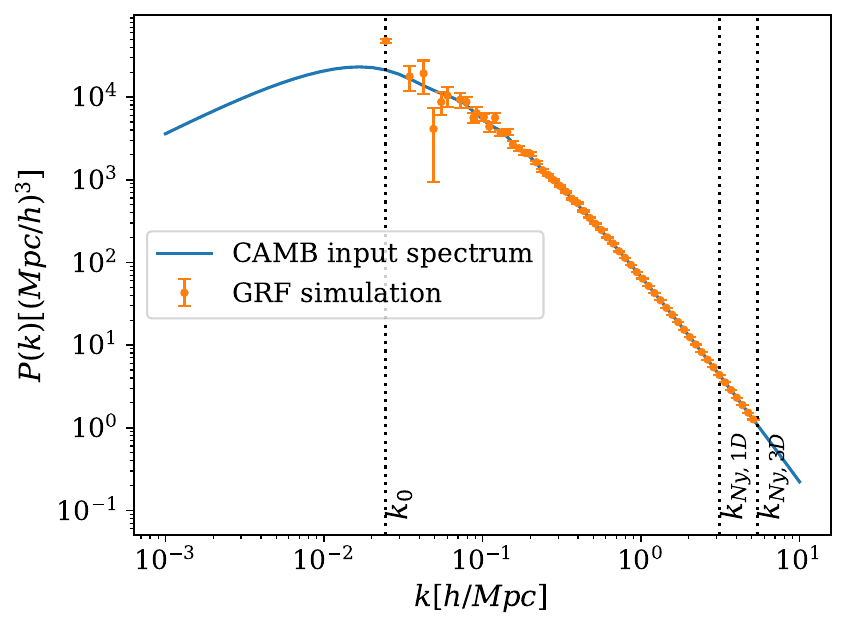}
  \caption{Left: Random realization of a Gaussian random field in a cubic periodic domain $\mathcal{D}=\mathbb{T}^3$. Right: Corresponding isotropically averaged power spectrum of the field realization along with the input expectation value obtained with \textsc{Camb}. The vertical lines indicate the fundamental mode of the box $2\pi/L$ and the 1D and 3D Nyquist mode $N \pi/L$, and $\sqrt{3} N\pi/L$.}
  \label{fig:GRF:sim}
\end{figure}

\noindent We can therefore generate a centered homogeneous and isotropic GRF by sampling in the diagonal Fourier eigenbasis. To make this computationally tractable in our case, we discretize the problem by imposing periodic boundary conditions on a three-dimensional box of unit side length, which mathematically corresponds to working on a unit 3-torus $\mathbb{T}^3=[0,1)^3$. On this periodic domain, any square-integrable function must satisfy $f(\vec{x}+\vec{m})=f(\vec{x})$ for all $\vec{m}\in\mathbb{Z}^3$. This periodicity constraint restricts the allowed Fourier modes to those that are also periodic, requiring the wave vectors to take the form $\vec{k}=2\pi \vec{m}$. Thus, the continuous Fourier integral reduces to a discrete Fourier series and the field can be written as
\begin{align}
  f(\vec{x}) = \text{Re}\;\sum_{\vec{m} \in \mathbb{Z}^3} \hat{f}_{\vec{m}} \,\ee^{\ii 2\pi \vec{m}\cdot \vec{x}}\;. 
\end{align}
We have to take the real part since, without additional constraints on the coefficients, the resulting field would in general be complex. Alternatively, one can require the coefficients to obey a Hermitian symmetry, i.e. one imposes by hand that $\hat{f}_{\vec{m}} = \hat{f}^\ast_{-\vec{m}}$ holds\footnote{This can also be exploited to reduce the memory footprint allowing the use of complex-to-real FFTs instead of complex-to-complex which require about twice the memory.}. In practice this sum can be efficiently computed using the Fast Fourier Transform (FFT) if one restricts to a finite number of $N^3$ modes, i.e. if $\vec{m}\in \left[-\tfrac{N}{2},\tfrac{N}{2}\right)^3 \subset \mathbb{Z}^3$. The modes of the highest frequency ($\pm N/2$) are called `Nyquist modes'. 

The discrete Fourier coefficients to be sampled and from which the discretized Gaussian field realization is finally constructed are therefore
\begin{align}
\hat{f}_{\vec{m}} = \sqrt{P(2\pi\|\vec{m}\|)} \;\frac{\alpha_{\vec{m}} + \ii \beta_{\vec{m}}}{\sqrt{2}}\qquad\text{with}\qquad \alpha_{\vec{m}},\beta_{\vec{m}}\sim\mathcal{N}(0,1),
\end{align}
and it is easy to verify that then indeed $\mathbb{E}[\hat{f}_{\vec{m}}\hat{f}^\ast_{\vec{m}}]  = P(2\pi\|\vec{m}\|)$. Here, $P(\cdot)$ is a given isotropic power spectrum, taken usually as the matter power spectrum obtained from an Einstein-Boltzmann code such as \textsc{Camb}\footnote{\textsc{Camb} is available from \url{https://camb.info}} or \textsc{Class}\footnote{\textsc{Class} is available from \url{http://class-code.net}}.  In particular, given the matter power spectrum $P_m$, a realization of the primordial density field $\delta(\vec{x})$ can be made in a few lines of code, as shown in \autoref{fig:GRF:sim}, you can try it out yourself \href{https://github.com/ohahn/LesHouches2025/blob/main/LSSfromGRF.ipynb}{in the accompanying notebook}. In practice, the co-moving box size $L$ has to be used to relate the discrete Fourier modes to the modes output by the Einstein-Boltzmann code.

\section{The dynamics of cold collisionless matter}
\subsection{Definitions, the cosmological Vlasov-Poisson system} 
On large enough scales and at late times, the evolution of cosmic matter on sub-horizon scales can be described as that of a cold collisionless fluid evolving under its Newtonian self-gravity. Assume the one-particle phase space is given by $\mathcal{P}:=\mathcal{D}\times\mathbb{R}^3\times[0,T]$, where $\mathcal{D}\subset\mathbb{R}^3$ is usually a 3-torus (periodic box) $\mathbb{T}^3$. For simplicity, we normalize the volume so that $\int_\mathcal{D}\dd^3 x=1$. The density in phase space of matter is given by a positive definite density $f(\vec{x},\vec{v},t)\ge 0$, and we want to describe the fluid in terms of the evolution of this density function. Such a system is described by the cosmological Vlasov-Poisson (VP) equations, see e.g. \cite{Peebles:1980:LSSbook,AH_review:2022,Rampf:2021:review} for more details.
\begin{subequations}
\begin{definition}[Cosmological VP system]
  In co-moving coordinates, the VP equations\footnote{Note that time units are in Hubble time (and hence $H_0=1$).} are given by 
  \begin{align}
  &\text{(V)}\quad&
      \frac{\partial f}{\partial t} + \frac{\vec{v}}{a^2}\cdot \bnabla_x f - \bnabla_x\phi\cdot\bnabla_v f &= 0 && \label{eq:Vlasov}\\ 
  &\text{(P)}\quad&
      \nabla_x^2 \phi + \frac{\kappa}{a} \left(1-n \right) &= 0 \label{eq:Poisson}&
    \quad \text{with } \quad n &:= \int_{\mathbb{R}^3} f\; \dd^3 v \quad\text{and}\quad \int_\mathcal{D} n\; \dd^3 x = 1
  \end{align}
  where $\phi$ is the gravitational potential, $\kappa := 3\Omega_\text{m}/2$ is a cosmology dependent constant, and $n$ is the configuration space density. The `scale factor' $a(t)$ is the solution to the Friedmann equation $\dot{a} = aH(a)$ where $H(a)$ is the Hubble function, and e.g. for a flat $\Lambda$CDM universe
  \begin{align}
    H(a) = \sqrt{\Omega_\text{m} a^{3} + (1-\Omega_\text{m})}\;.
  \end{align}
\end{definition}
  Since the fluid is cold, it has (initially) no extent in velocity space at fixed location, and we specify the initial data solely in terms of an initial velocity perturbation, i.e. at $t=0$, we set for a cold fluid
  \begin{align}
    f(\vec{x},\vec{v},t_0) = \delta_D( \vec{v} - \vec{v}_0(\vec{x}))\;. \label{eq:Vlasov:ICs}
  \end{align}
\end{subequations}
  Eqs.\eqref{eq:Vlasov}-\eqref{eq:Vlasov:ICs} constitute a rather non-trivial set of non-linear partial differential equations, and we typically have to resort to numerical or perturbative solutions to arrive at approximate solutions for general initial data. 

  \subsection{Solution by method of characteristics}
  The Vlasov equation \eqref{eq:Vlasov} can be solved by the method of characteristics. Characteristics are a family of one-parameter curves $\vec{X}(t),\vec{V}(t)$ in phase space that transport the initial data across space and time. They are defined as the solutions to the characteristic ordinary differential equations (ODEs). The total derivative of the phase space density $f$ along the characteristic curves is given by
  \begin{align}
    \frac{\dd }{\dd t}f(\vec{X}(t),\vec{V}(t),t) &= \frac{\partial f}{\partial t} + \dot{\vec{X}}\cdot \bnabla_x f + \dot{\vec{V}}\cdot\bnabla_v f \;.
  \end{align}
  Clearly, the phase space density $f$ is conserved along the characteristic curves, i.e. $\dd f/\dd t=0$, if the characteristic ODEs are chosen such that the derivatives match the respective terms of the Vlasov equation, i.e. if
  \begin{align}
    \dot{\vec{X}} &= \frac{\vec{V}}{a^2}\;, \qquad \dot{\vec{V}} = -(\bnabla_x \phi)(\vec{X})\;.\label{eq:charODE:cosmictime}
    \intertext{which is equivalent to a second order ODE}
    \ddot{\vec{X}} &= -2 H\dot{\vec{X}} - \frac{1}{a^2} (\bnabla_x \phi)(\vec{X})\;.
  \end{align}
  The solution is fully determined once the initial data \eqref{eq:Vlasov:ICs}, i.e. $\vec{X}(0)=\vec{q}$ and $\vec{V}(0)=\vec{v}_0(\vec{q})$ is specified for all $\vec{q}\in\mathcal{D}$. The density $n$ can be shown to be related to the conservation of the measure 
  \begin{align}
    n\; \dd^3 x = \dd^3 q\qquad \Leftrightarrow \qquad n(\vec{X},t) \;\left|\frac{\partial \vec{X}}{\partial \vec{q}}\right| = 1\;,
  \end{align}
  where we usually write the Jacobian determinant as $J$, i.e. $n = 1/|J|$, so that the formal solution to the Poisson equation \eqref{eq:Poisson} is given by
  \begin{align}
    \phi(\vec{X},t) &= \frac{\kappa}{a} \bnabla_X^{-2} \frac{J-1}{J} \;.
  \end{align} 
  We will leave this formal result as it is for now, and consider perturbative solutions later.
  \begin{definition}[Shell-crossing singularity]
    For a fluid with cold initial data, there can (and generally will) be a time $t_\ast$ when the mapping $\vec{q}\mapsto \vec{X}(\vec{q},t_\ast)$ becomes multivalued and thus no longer one-to-one. This is accompanied by a vanishing of the Jacobian $J$ of the map, leading to a formal divergence of the density $n=1/|J|$. Shell-crossing represents a fundamental breakdown of the single-stream approximation: when fluid elements following different trajectories arrive at the same spatial location, the velocity field becomes multi-valued. This is the regime where dark matter halos form.
  \end{definition}
  
  \subsection{Cosmological Euler-Poisson via the Boltzmann hierarchy} The integrating out of velocity degrees of freedom when going from the phase space density $f$ to the configuration space density $n$ can be applied systematically by performing an expansion of $f$ in terms of velocity moments of the phase space distribution function
  \begin{align}
    n(\vec{x},t) &:= \int_{\mathbb{R}^3} f(\vec{x},\vec{v},t)\; \dd^3 v\;; &
    n{\vec{u}} &:= \int \vec{v} f \dd^3 v\;; &
    n\left({\mat{\sigma}}+{\vec{u}}\otimes{\vec{u}}\right) &:= \int \vec{v}\otimes \vec{v} f \, \dd^3 v \;,\quad \ldots\;,
  \end{align}
  which correspond to density, momentum density, and total stress-energy density tensor, and we omitted the explicit function arguments for the second two expressions. The same expansion in terms of moments can be applied to the Vlasov equation, which yields an infinite hierarchy (the `Boltzmann hierarchy') of equations, the first two of which are the only ones we shall consider here. By taking velocity moments of the Vlasov equation, we systematically eliminate velocity-space dependence, trading a 6D phase-space description for a 3D configuration-space description with additional moment variables.
  
    \begin{subequations}
  \begin{definition}[Cosmological Euler-Poisson system]
    The first two marginals in the Boltzmann hierarchy correspond to the cosmological Euler-Poisson system (cf. \cite{Peebles:1980:LSSbook})
  \begin{align}
    \frac{\partial n}{\partial t}+\frac{1}{a^2}\bnabla_x \cdot \left(n{\vec{u}}\right) &=0  \label{eq:Continuity}\\
    \frac{\partial {\vec{u}}}{\partial t} +\frac{1}{a^2}\left({\vec{u}}\cdot\bnabla_x\right)  {\vec{u}}  &= -\bnabla_x\phi - \frac{1}{n a^2}\bnabla_x\cdot\left(n{\mat{\sigma}}\right) \label{eq:Euler}
  \end{align}
  supplemented with the Poisson equation \eqref{eq:Poisson} from above. 
\end{definition}
  With initial data given by \eqref{eq:Vlasov:ICs}, the respective initial data for the hierarchy becomes 
  \begin{align}
    n(\vec{x},0) &= 1, & \vec{u}(\vec{x},0) &= \vec{v}_0(\vec{x}), & \mat{\sigma}(\vec{x},0) &= 0\;.
  \end{align}
\end{subequations}
  In the cold limit, one thus has ${\mat{\sigma}}=0$ initially, and the system is closed. However, even if $\mat{\sigma}=0$ initially, internal anisotropic stress will be generated (non-perturbatively) during shell-crossing. Before shell-crossing, the term $\mat{\sigma}$ can however be assumed to vanish everywhere and thus dropped from the equation.

  You have learned in the introductory lectures that equations \eqref{eq:Continuity}-\eqref{eq:Euler} plus the Poisson equation can be solved perturbatively by defining $\delta := n-1$ as a small parameter. At first order (of the standard perturbation theory) one has the well known result
  \begin{align}
    \delta(\vec{x},t) &= D_+(t) \,\delta_+(\vec{x}) + D_-(t) \,\delta_-(\vec{x})\;,
  \end{align}
i.e. the linear solution separates into a growing and a decaying mode, where $D=D_\pm$ solve the so-called linear `Eulerian perturbation equation'
\begin{align}
  \ddot{D} +2H \dot{D} - \frac{3}{2}H_0^2\Omega_\text{m} \frac{D}{a^3}=0. \label{eq:euler_pt_eq}
\end{align}
and the spatial pieces $\delta_\pm(\vec{x})$ are determined by the initial condition. In the case of a flat $\Lambda$CDM universe, one has \cite{Chernin:2003}
\begin{align}
 D_+ &= a  \,\sqrt{1+ \lambda_0 \,a^3}\,\, {}_2F_1 \left( \frac 3 2, \frac 5 6, \frac{11}{6}, - \lambda_0\, a^3\right) ,  \label{eq:D+} &
D_-  &= a^{-3/2} \;\sqrt{1+ \lambda_0\, a^3}, 
\end{align}
where ${}_2F_1$ is Gauss' hypergeometric function, and $\lambda_0 := (1-\Omega_\text{m}) / \Omega_\text{m}$. In Newtonian gravity, for a cold fluid, linear growth is thus completely scale-independent. This is of course not true in relativistic perturbation theory, since the \textit{horizon scale} introduces a physical scale with sub- and super-horizon scales obeying different growth rates. In the presence of finite temperature effects, another scale, the \textit{Jeans scale}, enters below which the growth of perturbations is suppressed.

\subsection{Vlasov early time asymptotics: the Zel'dovich approximation}
A particularly useful result can be obtained by expressing the Vlasov equation \eqref{eq:Vlasov} using the growth function $D_+$ as the time variable. This works of course as long as there is a monotonous relation between $t$ and $D_+$. By choosing the growth factor $D_+$ as the time variable rather than cosmic time $t$, we can isolate the time-dependence and reveal the asymptotic structure of the solution more clearly. Defining a new, re-scaled, velocity $\vec{w}:= \frac{\vec{v}}{a^2\dot{D}_+}$, one can re-write the Vlasov-Poisson system for $a\to0$ as (using $D:=D_+$)
\begin{align}
  \frac{\partial f}{\partial D}  + \vec{w}\cdot \bnabla_x f -\frac{\kappa D}{a^3\dot{D}^2}\left(  \vec{w} + \bnabla_x \varphi\right)\cdot\bnabla_w f &= 0\;. &
  \nabla^2_x \varphi - \frac{\delta}{D} &= 0 
\end{align}
where the potentials are related as $\varphi = \frac{a}{\kappa D}\phi$. The characteristic ODEs can again be combined into a single second order ODE, defining the growth rate $f_g := \dd \log D/\dd \log a$,
\begin{align}
  \vec{X}' &= \vec{W}\;, & \vec{W}' &= -\frac{\kappa D}{a^3\dot{D}^2}\left( \vec{W} + \bnabla_X \varphi\right)\;,\\
  \Leftrightarrow\qquad\vec{X}'' &= -\frac{\kappa}{a^3f_g^2 H^2 D}\left( \vec{X}' + \bnabla_X \varphi\right)\;.\label{eq:lpt:pre-master}
\end{align}
where a prime now denotes the derivative with respect to $D$ and $\bnabla_X\varphi=(\bnabla_x \varphi)(\vec{X})$. To find an asymptotic solution for early times as $a\to 0$, we can use the fact that the growth factor $D$ is given by\footnote{For two functions $g(x), h(x)$, we write $g \asymp h$ as $x\to x_0$ to denote that $g$ and $h$ are asymptotically equivalent, i.e., $\lim_{x\to x_0} g(x)/h(x) = 1$.} $D \asymp a$ in the matter dominated era of $\Lambda$CDM (one neglects here the radiation dominated phase). The whole prefactor has the following early time asymptotic behavior, allowing to determine the asymptotic form of the characteristic ODE 
\begin{align}
  \frac{\kappa}{a^3f_g^2 H^2 D} &\asymp  \frac{3}{2a} & \Rightarrow& &
   \vec{X}' + \bnabla_X \varphi &= \frac{2a}{3}\vec{X}'' \stackrel{a\to 0}{\to} 0\;.
\end{align}
Therefore, asymptotically, the characteristics are given by the asymptotic solution for $a\to 0$ in terms of the initial velocity perturbation $\vec{W}_0 = -\bnabla_q\varphi_0$ as \cite{Brenier:2003}
\begin{align}
  \vec{X}_0 &= \vec{q} &
  \vec{X}'_0 &= - \bnabla_q \varphi_0(\vec{q}) & \Rightarrow\qquad \vec{X} &= \vec{q} - D_+ \,\;\bnabla_q \varphi_0(\vec{q}) + \mathcal{O}(D_+^2)\;,
\end{align}
This is the Zel'dovich approximation (ZA, \cite{Zeldovich:1970}): at leading order, characteristics are straight lines in $D_+$-time (only the $D_+$ branch is finite as $a\to0$) with the slope of the line set by $\bnabla_q \varphi_0$.

\subsection{Lagrangian perturbation theory}
We first write the characteristic solution for the position in terms of the initial (so-called Lagrangian) coordinate and a displacement relative to it as 
\begin{align}
  \vec{X}(\vec{q},t) = \vec{q} + \vec{\Psi}(\vec{q},t)
\end{align}
The Jacobian matrix $\mat{J}$ of this coordinate change is in index notation given by 
\begin{align}
   (\mat{J})_{ij} &= \frac{\partial X_i}{\partial q_j} = \delta_{ij} + \Psi_{i,j} & &\Rightarrow &
 \bnabla_X &= \mat{J}^{-1} \cdot \bnabla_q & &\text{and}  & J = \det \mat{J}\;.
\end{align}
We first write eq.~\eqref{eq:lpt:pre-master} in slightly more concise form inserting the Poisson equation as 
\begin{align}
  \mathfrak{D} \vec{X} &= \bnabla_X \nabla_X^{-2}\left(1 - \frac{1}{J}\right) & &\text{where} & \mathfrak{D} := \frac{a^3}{\kappa}\left(\frac{\dd^2}{\dd t^2} + 2H \frac{\dd }{\dd t}\right)
\end{align}
is a differential operator acting only on time. Next, we hit this equation with another $\bnabla_X$ to decompose it into its longitudinal and transversal parts, i.e. in index notation
\begin{align}
  \frac{\partial q_j}{\partial X_i }\mathfrak{D} \frac{\partial X_i}{\partial q_j} &= 1- \frac{1}{J} & &\text{and} & \epsilon_{ijk}\frac{\partial q_l}{\partial X_j }\mathfrak{D}\frac{\partial X_k}{\partial q_l} &= 0 \label{eq:LPT:master}\;.
\end{align}
The second equation expresses the conservation of vorticity, but since we must assume that the vorticity is zero initially (vorticity is a decaying mode), we can directly impose the stronger constraint that $\bnabla_X \times \frac{\dd \vec{X}}{\dd t}=0$. Here, we will neglect the transversal part entirely as it only enters at third order of the perturbation theory and focus on the longitudinal modes exclusively.

We now make a weak perturbative ansatz of the form
\begin{subequations}
\begin{align}
  \vec{X}& = \vec{q} + \epsilon\; D^{(1)}(t) \;\bnabla_q {\phi}^{(1)}  + \epsilon^2 \;D^{(2)}(t) \;\bnabla_q{\phi}^{(2)} + \ldots\\
  \Rightarrow \qquad \mat{J} &= \mat{I} + \epsilon \mat{A} \qquad \text{with}\quad (\mat{A})_{ij} = D^{(1)}(t) {\phi}^{(1)}_{,ij}  + \epsilon D^{(2)}(t) {\phi}^{(2)}_{,ij} + \ldots
\end{align}
In three dimensions, the following formula holds for the determinant of a $3\times3$ matrix $\mat{I} + \epsilon \mat{A}$:
\begin{align}
  J = \det \mat{J} &= \det(\mat{I}+\epsilon \mat{A}) = 1 + \epsilon \, \text{Tr}(\mat{A}) + \frac{\epsilon^2}{2} \left(  \text{Tr}(\mat{A})^2 - \text{Tr}(\mat{A}^2) \right) + \epsilon^3 \det \mat{A}\;,\nonumber\\
  &=: 1+\epsilon \;\mu_1(\mat{A}) + \epsilon^2 \;\mu_2(\mat{A}) + \epsilon^3\; \mu_3(\mat{A}) \;.
  \intertext{and therefore}
  1-\frac{1}{J} &= \epsilon \, \mu_1 - \epsilon^2\left(\mu_1^2-\mu_2\right) + \epsilon^3\left(\mu_1^3-2\mu_1\mu_2+\mu_3\right) + O(\epsilon^4)\;.
\end{align}
The von Neumann series of the inverse of a matrix allows us to write
\begin{align}
  \mat{J}^{-1} = (\mat{I} + \epsilon \mat{A})^{-1} &= \sum_{n=0}^\infty (-\epsilon)^n \mat{A}^n = \mat{I} - \epsilon \mat{A} + \epsilon^2 \mat{A}^2 - \ldots\;.
\end{align}
\end{subequations}
\paragraph{First order solution (1LPT=ZA).} Keeping all terms to first order in $\epsilon$, we find 
\begin{align}
  \left(\mathfrak{D}-1\right) D^{(1)} \phi^{(1)}_{,ii} &= 0 & &\Rightarrow & \left(\mathfrak{D}-1\right) D^{(1)}&=0
\end{align}
 which is (unsurprisingly) the same as the linear Eulerian PT eq.~\eqref{eq:euler_pt_eq}, so $D^{(1)}=D_+$ and $\phi^{(1)}_{,ii} = \delta_+$. And with initial data $\phi^{(1)}= -\varphi^\text{ini}$ this yields of course the Zel'dovich approximation
\begin{align}
  \vec{X} = \vec{q} - D_+ \bnabla_q \varphi^\text{ini}. \label{eq:Zeldovich_solution}
\end{align}
Note that if the initial data $\varphi^\text{ini}$ is planar one-dimensional, then all higher order LPT terms vanish, and the Zel'dovich approximation is actually exact (until shell-crossing).

\paragraph{Second order solution (2LPT).} Keeping terms of order $\epsilon^2$ we find (since $\mathfrak{D}D^{(1)}=D^{(1)}$)
\begin{align}
  \mathfrak{D}D^{(2)} \phi_{,ii}^{(2)} &= -D_+^2 \;\mu_2\bigl( \phi^{(1)}_{,ij}\bigr)\;.
\end{align}
Equating the temporal and the spatial pieces separately, one has \cite{Buchert:1993,Bouchet:1995}
\begin{align}
  \left(\mathfrak{D}-1\right)D^{(2)} &= -D_+^2 & &\text{and} & \phi^{(2)} =  \nabla_q^{-2}\tfrac{1}{2}\left(\phi^{(1)}_{,ij}\phi^{(1)}_{,ij}-\phi^{(1)}_{,ii}\phi^{(1)}_{,jj}\right)
\end{align}
The source term $\mu_2$ encodes leading-order tidal effects. The ODE for the second order growth function $D^{(2)}$ has to be integrated numerically usually. If few per cent level accuracy is enough, the following approximate form can however be used \cite{Bouchet:1995}
\begin{align}
  D^{(2)} \approx -\frac{3}{7} \Omega_\text{m}^{-2/63} D_+^2\;.
\end{align}

\begin{figure}[ht]
  \centering
  \includegraphics[width=0.46\columnwidth,valign=c]{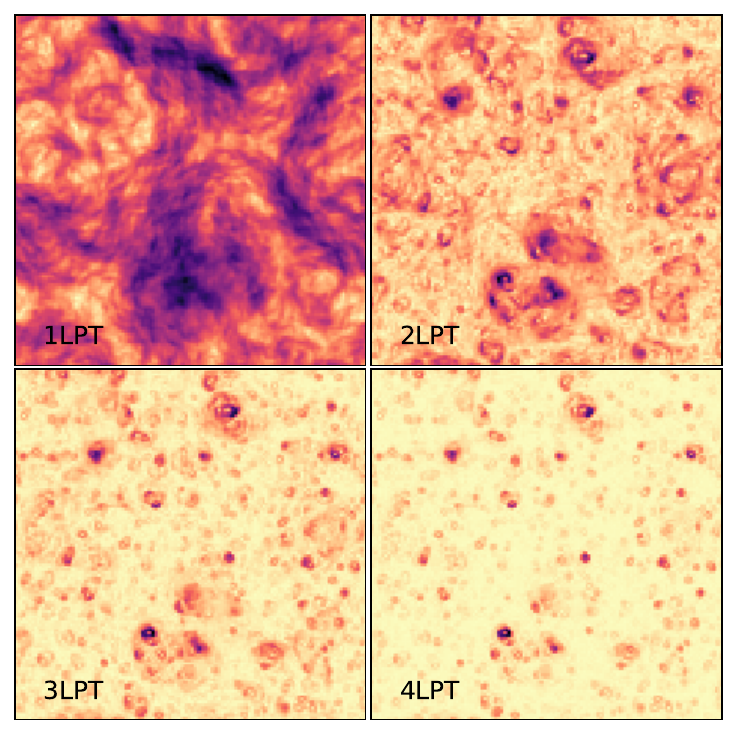}\qquad
  \includegraphics[width=0.45\columnwidth,valign=c]{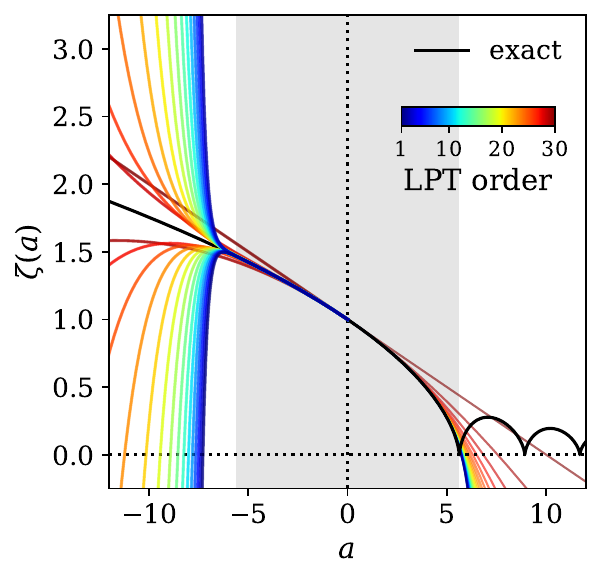}
  \caption{Left: Norm of the displacement field $\|\vec{\psi}^{(n)}\|$ as a function of the initial Lagrangian coordinate $\vec{q}$ up to 4LPT. A darker color corresponds to a larger displacement, each panel is individually normalised. The slowest converging regions are the almost spherically collapsing and expanding regions. (Adapted from \cite{Rampf:2021}) Right: Convergence of the radius $\zeta(a)$ of a spherical perturbation in LPT, where $\zeta(a)=1+\sum_{n=1}^N a^n \psi^{(n)}$ with $n$-th order displacement $\psi^{(n)}$. The plot shows the convergence from 1LPT ($N=1$, dark red line) to 30LPT ($N=30$, dark blue line) to the trajectory of spherical collapse (black line). The radius of convergence in both the over- and underdense case is determined by the shell-crossing singularity at $a_\ast=\tfrac{1}{2}(12\pi)^{2/3}\approx5.6$ which limits the convergent regime to the gray shaded area.}
  \label{fig:displacement_norm}
\end{figure}

\paragraph{All-order recurrence relations (nLPT).} It is possible to determine all order recurrence relations for LPT without too much further effort from the master equations \eqref{eq:LPT:master} \cite{Rampf:2012,Zheligovsky:2014,Matsubara:2015}. These allow to compute the $n+1$-th order based on the previous orders up to $n$. In the left panel of \autoref{fig:displacement_norm}, the contributions $\|\psi^{(n)}\|$ for GRF initial conditions are shown (see also \cite{Rampf:2021}). The slowest convergent pieces are close to spherically expanding/collapsing regions. When computing the nLPT series for spherical collapse (see lectures by Cora Uhlemann), one sees that even in the underdense case, the convergence of the nLPT perturbative expansion is limited by the shell-crossing singularity which defines the convergence radius. This means that we can safely approach the regime just before shell-crossing singularities appear using high order LPT, but have to resort to $N$-body simulations to go beyond shell-crossing. Equations up to 3LPT can be found e.g. in \cite{Michaux:2021}, pseudo code for $n$LPT recurrence in the appendix of \cite{Rampf:2021}.

\section{$N$-body simulations and PT-informed time integration}
As discussed above, $n$LPT is exceedingly accurate in the regime before shell-crossing singularities appear. To go beyond, one has to resort to $N$-body simulations.

\subsection{$N$-body initial conditions}
Ever since the early days of cosmological $N$-body simulations in the 1980s, LPT has been used to initialize $N$-body simulations \cite{Efstathiou:1985}. Initially using 1LPT/ZA, increasing precision of simulations in the early 2000s \cite{Crocce:2006} indicated that there can be `transients' appearing when 1LPT is used. These are effectively truncation errors due to neglecting higher order LPT terms in the initial conditions. The simulation has to evolve for some time for the truncation errors to become small compared to the growth of nonlinearities in the simulation itself -- this is particularly true for higher order statistics such as 3-point functions and higher. More recently \cite{Michaux:2021} pointed out that for current precision requirements, it is best to use high-order LPT and an as late as possible time to start the $N$-body simulation to reduce both truncation errors and discreteness effects (for the latter see \autoref{sec:discreteness}). 

There is a mismatch in physics between what we include in the Einstein-Boltzmann solvers (linear but fully relativistic) and what we typically include in simulations of the late Universe (non-linear but non-relativistic). This is justified as we have a good separation of scales (in time and space) where these two are relevant: nonlinearities are relevant only on scales much smaller than the horizon. This means that horizon-scale effects can be safely modeled using linear corrections if needed, and other post-Newtonian effects on small cosmological scales are of order $\sim 10^{-5}$ \cite{Adamek:2016}. For all simulations that cover scales smaller than the horizon, relativistic effects can thus be safely neglected. 

The steps to initialize an $N$-body simulation are thus
\begin{enumerate}
  \item Compute a matter power spectrum $P_m(k)$ at a target redshift, e.g. $z_\text{target}=0$ using an Einstein-Boltzmann code (e.g. \textsc{Camb} or \textsc{Class} in synchronous gauge).
  \item Simulate a primordial potential $\varphi_0$ as a GRF with a spectrum $P_\varphi = k^{-4}P_m(k) / D_+(z_\text{target})^2$.
  \item Compute the nLPT spatial terms and assemble the total displacement field to $n$-th order for each particle for growth factors $D^{(1)}, D^{(2)},\dots$ at $z_\text{start}$
  \begin{align}
    \vec{X}_i &= \vec{q}_i + D^{(1)} \;\vec{\Psi}^{(1)}(\vec{q}) + D^{(2)} \;\vec{\Psi}^{(2)}(\vec{q}) + \ldots
    \intertext{determine the initial particle velocity (careful about definition, which may be different for different $N$-body codes)}
    \vec{V}_i = \frac{\dd \vec{X}_i}{\dd t} &= \dot{D}^{(1)} \;\vec{\Psi}^{(1)}(\vec{q}) + \dot{D}^{(2)}\; \vec{\Psi}^{(2)}(\vec{q}) + \ldots
  \end{align}
  This is most easily achieved by placing the $\vec{q}_i$ on a simple cubic lattice. 
  \item Start your $N$-body simulation at $z_\text{start}$ and evolve to whatever time you are interested in using time integration as explained next, and using Poisson solvers as explained in Romain Teyssier's lecture.
\end{enumerate}
An example of how to carry out steps 1 to 3 can be found in a \href{https://github.com/ohahn/LesHouches2025/tree/main}{\textsc{Jupyter} notebook accompanying these notes}. Software that implements these steps includes e.g. \textsc{2lptIc}/\textsc{N-GenIC}\footnote{The original \textsc{2lptIc} is available from \url{https://cosmo.nyu.edu/roman/2LPT/}, and \textsc{N-GenIC} which builds on it from \url{https://www.h-its.org/2014/11/05/ngenic-code/}} for 2LPT ICs, and \textsc{MonofonIC}\footnote{\textsc{MonofonIC} is available from \url{https://github.com/cosmo-sims/monofonIC}} for 3LPT ICs and for multi-fluid baryon+CDM ICs.

\subsection{Standard time integrators}
Based on the characteristic equations in cosmic time eqs.~\eqref{eq:charODE:cosmictime}, it is natural to advance particles using a leapfrog scheme that advances (e.g. in drift-kick-drift form) \cite{Quinn:1997} with $a_{n+1}-a_n = \Delta a$ to evolve positions and velocities for particle $i=1,\ldots,N$ by one time step 
\begin{subequations}
 \begin{align}
    \vec{X}^{n+1/2}_i &= \vec{X}^n_i + \vec{V}^n_i \int_{a_n}^{a_{n+1/2}} \frac{\dd a}{a^3H}\;,\\
     \vec{V}^{n+1}_i &= \vec{V}^{n}_i -\bnabla \Phi^{n+1/2} \int_{a_n}^{a_{n+1}} \frac{\dd a}{a^2 H}\;,\\
     \vec{X}^{n+1}_i &= \vec{X}^{n+1/2}_i + \vec{V}^{n+1}_i \int_{a_{n+1/2}}^{a_{n+1}} \frac{\dd a}{a^3H}\;.
  \end{align}
\end{subequations}
This integrator can be proven to be second order accurate and symplectic (see \cite{Hairer:2006} for definitions and properties of geometric integrators). It is however non-optimal for cosmological simulations, as it can be shown to not reproduce the Zel'dovich solution for one-dimensional initial data exactly (even though that just consists of inertial motion!). The reason is that it uses poorly chosen time and velocity definitions for large-scale evolution. This can be formally quantified as follows.

\begin{definition}[Zeldovich consistency]
    A time integrator is Zel'dovich consistent, if it reproduces the Zel'dovich solution \eqref{eq:Zeldovich_solution} for one-dimensional initial data exactly in a single time step. See \cite{List:2024} for more details.
\end{definition}
\subsection{PT-informed integrators}
Instead of using the cosmic time velocity, we can employ the velocity we defined above that lead to the inertial Zel'dovich motion asymptotically, i.e. we define the velocity as 
\begin{align}
  \vec{W} &:= \vec{V}/F &\text{with}\qquad F&:=a^2\dot{D}
\intertext{so that}
  \vec{X}' &= \vec{W}\;, & \vec{W}' &= -\frac{\kappa D}{a^3\dot{D}^2}\left( \vec{W} + \bnabla_X \varphi\right)\;,
\end{align}
Let us formulate a more general $D$-time integrator (again in drift-kick-drift form), where now one steps in $D$, i.e. $D_{n+1}-D_n=\Delta D$, as
\begin{subequations}
\label{eq:generalized_leapfrog}
\begin{align}
  \vec{X}^{n+1/2}_i &= \vec{X}^{n}_i + \frac{\Delta D}{2} \vec{W}^n_i \\
  \vec{W}^{n+1}_i &= \alpha \vec{W}^{n}_i + \frac{\beta}{D_{n+1/2}}  \bnabla\varphi(\vec{X}^{n+1/2}_i)\\
  \vec{X}^{n+1}_i &= \vec{X}^{n+1/2}_i + \frac{\Delta D}{2} \vec{W}^{n+1}_i
\end{align}
\end{subequations}

\begin{theorem}[Zel'dovich consistent time integrator]
    The integrator \eqref{eq:generalized_leapfrog} is Zel'dovich consistent if the functions $\alpha$ and $\beta$ are chosen such that
    \begin{align}
        \beta &= 1-\alpha\;.
    \end{align}
    The proof is straightforward and can be found in \cite{List:2024}.
\end{theorem}

\noindent Multiple choices are thus possible as the class of Zel'dovich consistent integrators of the above form still allows the freedom to choose the function $\alpha$.
\begin{description}
  \item[Fast-PM]  For the `\textsc{Fast-PM}' code, the first PT-informed integrator was proposed by \cite{Feng:2016}. Although it was not proven in the original paper, this integrator is  symplectic and 2nd order, and thus the only flavor in this family that is both symplectic and Zel'dovich consistent. 
For the integrator to be symplectic, the function $\alpha$ must be chosen to recover the `canonical momentum' $\vec{V}_i$ at the end of the time step, i.e. 
    \begin{align}
        \alpha &= \frac{F_n}{F_{n+1}}\;.
    \end{align}
    \item[Bullfrog] The \textsc{Bullfrog} integrator \cite{Rampf:2025} abandons symplecticity but matches the trajectory to 2LPT. In order to achieve this, the coefficient $\alpha$ must be chosen as 
    \begin{align}
    \alpha &= \frac{E_{n+1}'-G_{n+1/2}}{E_{n}'-G_{n+1/2}} & \text{with}\qquad G_{n+1/2} &:= D_n^{-1} \left(E_n+ E_{n}'\frac{\Delta D}{2}\right) - D_{n+1/2}\;,
    \end{align}
    where $E:=D^{(2)}$ is the second order growth factor and $E'=\dd E/\dd D$. This integrator produces the most accurate non-linear evolution on large scales with few time steps.
\end{description}
Clearly, in the limit of many time steps, all integrators should converge to the same solution. That solution will however be impacted by the quality of the force calculation (see lecture notes by Romain Teyssier).

\subsection{Discreteness effects in cosmological simulations}
\label{sec:discreteness}
\begin{figure}[ht]
  \centering
  \includegraphics[width=0.51\columnwidth,valign=c]{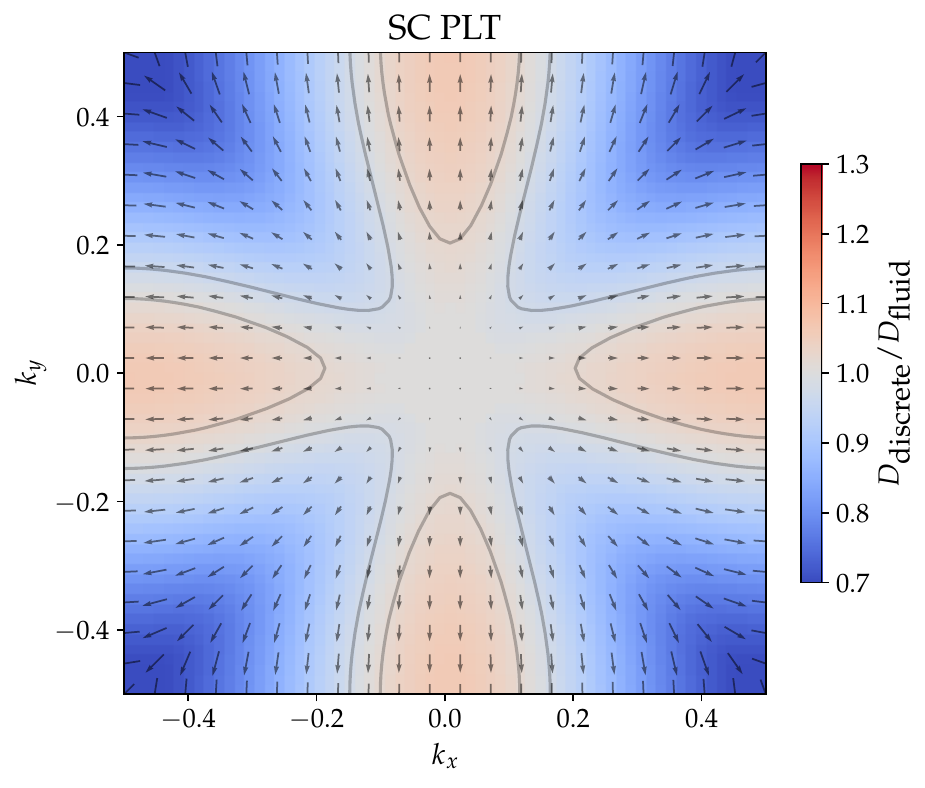}
  \includegraphics[width=0.48\columnwidth,valign=c]{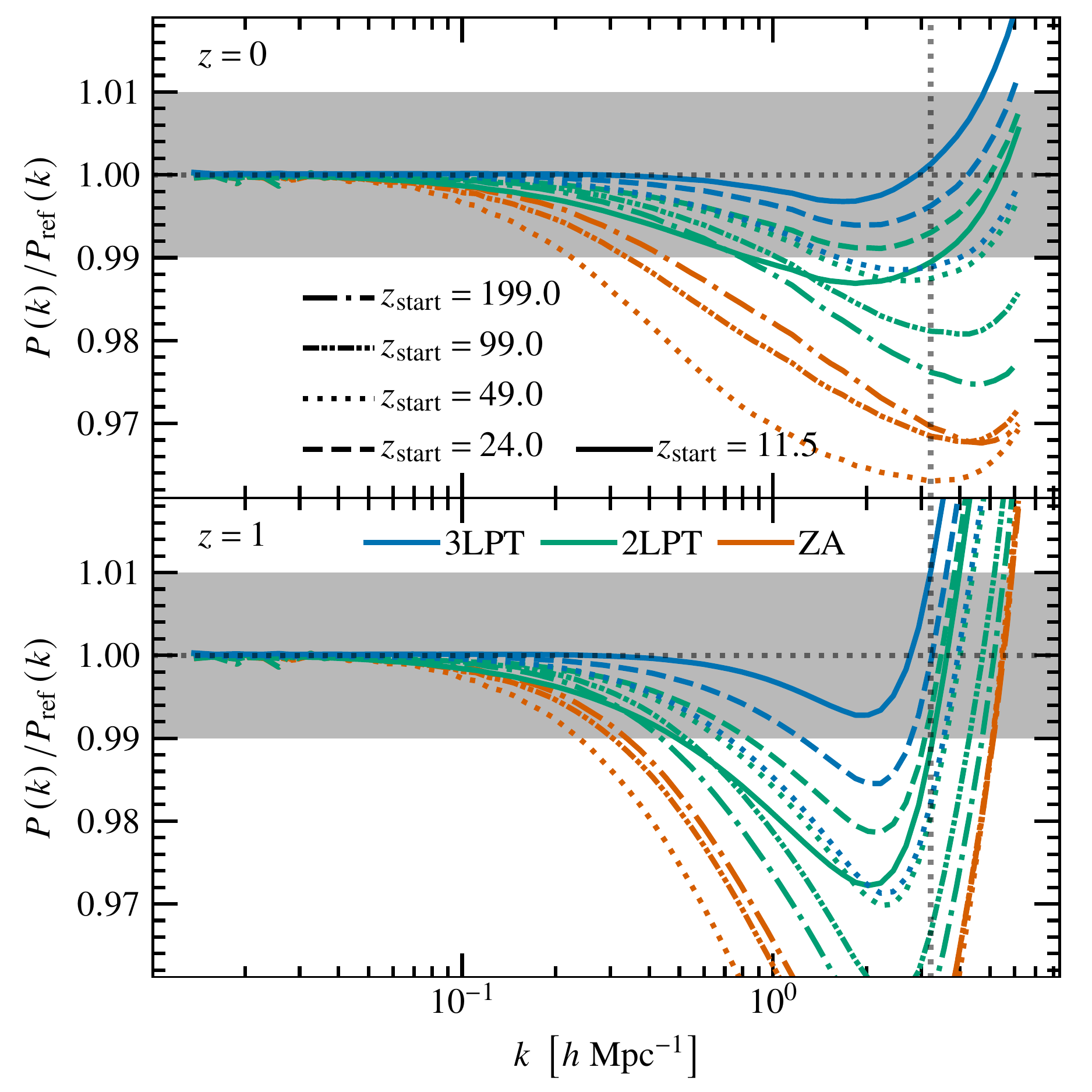}
  \caption{Left: Ratio of the growth factor of the discrete lattice system of particles to that of the fluid system in Fourier space ($k_{x,y}=\pm\tfrac{1}{2}$ corresponding to the Nyquist frequency). Arrows indicate the discrete growing mode eigenvectors, for the fluid limit they would be proportional to $\vec{k}$. One clearly sees that the highest wave numbers grow anisotropically and on average slower than in the fluid limit. Right: Error in the power spectrum at $z=0$ (top) and $z=1$ (bottom) depending on the starting redshift of the $N$-body simulation and the LPT-order used to initialize it (adapted from \cite{Michaux:2021}).}
  \label{fig:growth_factor}

\end{figure}

In these notes, we have derived two descriptions of non-linear structure formation as described by the Vlasov-Poisson system: the LPT perturbative approach, and the $N$-body simulation. The main difference is that in the $N$-body simulation, the Poisson source is approximated through the discrete characteristics, and re-expanded in each time step, while in nLPT the continuous evolution is computed order by order at the initial time. For the Poisson source one thus has
\begin{subequations}
\begin{align}
  1+\delta_\text{N-body}(\vec{x},t) &= \frac{1}{N}\sum_{i=1}^N \delta_D(\vec{x}-\vec{X}_i(t)) \\
  1+\delta_\text{LPT}(\vec{x},t) &= J^{-1}
\end{align}
\end{subequations}
It can be shown that at the particle scale, the $N$-body discretization leads to anisotropic deviations from the linear growth $D_+$. For particles starting from a simple cubic lattice, e.g., this deviation can be calculated exactly \cite{Joyce:2005}, and the ratio is shown in \autoref{fig:growth_factor} (left panel).

There are thus two competing numerical errors related to initial condition generation:
\begin{description}
  \item[LPT truncation error] As we are not using $\infty$LPT, there will be an error due to the truncation of the LPT expansion at finite order. This effect grows with time until shell-crossing appears and LPT becomes invalid. The error can be controlled by either starting the simulation at an earlier time (but see below), or by going to higher order LPT \cite{Michaux:2021}.
  \item[Discreteness errors]  In $N$-body simulations, the discrete particle system grows with a slightly modified growth rate compared to the continuous fluid (see \autoref{fig:growth_factor}, left panel). When simulations are initialized too early (i.e., when perturbations are still very small and linear), this discrete growth error accumulates over many time steps during the linear evolution phase. Since the error scales with the amplitude of fluctuations, its relative impact is largest when perturbations are small. These accumulated errors then become `frozen in' as the system becomes non-linear, degrading the final power spectrum on small scales. This error can be controlled by starting the `discrete' evolution as late as possible (when perturbations are already larger) \cite{Michaux:2021}, or by explicitly correcting for the modified discrete growth rate \cite{Garrison:2016}.
\end{description}
The impact on the $z=0$ and $z=1$ non-linear power spectrum of these two competing effects can be seen in \autoref{fig:growth_factor} (right panel). The conclusion made in \cite{Michaux:2021} from this analysis is that simulations are most accurately initialized with high order LPT at very late times, e.g. 3LPT at $z=24$ in their analysis -- which is quite in contrast to what was done before, where the common lore was that simulations should be initialized as early as possible.

\section{Conclusion}
In these lecture notes, we derived and discussed the main concepts and key aspects involved in generating initial conditions for cosmological $N$-body simulations from Gaussian random fields. The main takeaways are:
\begin{enumerate}
  \item Stationary GRFs have diagonal covariance in Fourier space, with expectation value given by the power spectrum $P(k)$. This property allows very efficient simulation of GRFs on periodic domains with a given $P(k)$ using discrete Fourier transforms.
  \item The Vlasov-Poisson system can be re-written in terms of ODEs for characteristic curves that preserve the phase space density. In LPT, the equations of motion of these characteristics are expanded perturbatively (in time), and can be used to evolve the system in the continuum limit. 
  \item In first order LPT, characteristics correspond to straight lines parameterized by the linear growth $D_+$ (this is the Zel'dovich approximation).
  \item Higher order LPT can be constructed by building on top of first order. LPT is convergent and valid before shell-crossing singularities arise (i.e. before characteristic curves intersect in position space).
  \item LPT displacements and velocities can be used as the starting point for $N$-body simulations, which can evolve the discrete system into the non-perturbative regime. 
  \item $N$-body simulations can use time integrators that respect LPT evolution. This enables them to converge on large scales with fewer time steps than when standard integrators are used.
  \item $N$-body simulations suffer from discreteness errors (due to the discrete particles approximating a continuous fluid). The effect of these errors can be controlled by starting the simulations as late as possible, from LPT of sufficiently high order.
\end{enumerate}

\begin{appendix}

  \numberwithin{equation}{section}

\bibliography{bibliography.bib}

\end{appendix}

\end{document}